\documentclass[twocolumn]{aastex631}

\usepackage{bm}
\usepackage{xcolor}
\usepackage{amsmath}
\usepackage{times}
\usepackage{accents}

\begin{document}
\title{Sample Variance in Cosmological Observations with a Narrow Field-of-View}
\shorttitle{Sample Variance in Cosmological Observations with a Narrow Field-of-View}
\shortauthors{Espenshade \& Yoo}
\author[0000-0002-9293-311X]{Peter Espenshade}
\affiliation{Center for Theoretical Astrophysics and Cosmology, Institute for Computational Science, University of Zürich, Winterthurerstrasse 190, CH-8057, Zürich, Switzerland}
\author[0000-0003-4988-8787]{Jaiyul Yoo}
\affiliation{Center for Theoretical Astrophysics and Cosmology, Institute for Computational Science, University of Zürich, Winterthurerstrasse 190, CH-8057, Zürich, Switzerland}
\affiliation{Physics Institute, University of Zürich, Winterthurerstrasse 190, CH-8057, Zürich, Switzerland}

\begin{abstract}
\noindent
    Surveys with a narrow field-of-view can play an important role in probing cosmology, but inferences from these surveys suffer from large sample variance, arising from random fluctuations around the cosmic mean. The standard method for computing the sample variance is based on two key approximations: treating perturbations linearly and the survey geometry as a box. We demonstrate that it can lead to a significant underestimate of the sample variance in narrow surveys. We present a new method for accurately computing the sample variance and apply our method to the recent observations of the warm-hot intergalactic medium (WHIM) based on spectroscopic measurements of blazars. We find that the sample variances in these surveys are significantly larger than the quoted measurement errors; for example, the cosmic mean baryon density contained in the WHIM could be lower by $54\%$ at $1\text{-}\sigma$ fluctuation than estimated in one observation. Accurately quantifying the sample variance is essential in deriving correct interpretations of the measurements in surveys with a small field-of-view.
\end{abstract}
    
\section{Introduction}
Large-scale surveys aim to derive cosmological parameters that determine the evolution of the Universe, and a larger survey volume is naturally preferred to obtain measurements of a better representative sample of the Universe (see, e.g., \citealt{SDSS, 2dFGRS, DES, BOSS, DESI}). In particular, since the initial conditions were given as a random realization around the cosmic mean value, there exist fluctuations in any given sample of finite volume (see, e.g., \citealt{PeeblesLSS_1980, Peacock_1998, Weinberg_2008, DodelsonSchmidt_2020}). The sample variance measures the dispersion of the sample volumes around the cosmic mean, and it is important to take into account the sample variance in deriving cosmological interpretations from a survey (\citealt{FKP_1994, MeiksinWhite_1999}). The sample variance always exists, and when the survey volume covers the whole observable Universe, the sample variance in this case is often referred to as the cosmic variance \citep{Hu_2003}. Naturally, the sample variance frequently dominates the error budget in a survey with a small volume like a pencil-beam survey (see, e.g., \citealt{KaiserPeacock_1991}).

In contrast, astronomical observations often target individual objects, and a preference in observational strategy is naturally given to a better spatial resolution than to a larger field-of-view. Care must be taken, however, if the goal of such astronomical observations with a small field-of-view is to derive cosmological interpretations or cosmic mean values. There exist three methods extensively used in the literature for computing the sample variance of a given cosmological survey. First, numerical simulations can provide mock samples of the Universe with precise survey geometries (e.g., \citealt{Norberg2009}), but they are costly to implement and depend on how baryon physics or the halo occupation model is adopted (e.g., \citealt{BerlindWeinberg_2002, Giri_2021}). Second, using sub-samples to estimate the sample variance is usually possible for large-scale surveys (e.g., \citealt{HillDriver2010, driver}), but for surveys with a narrow field-of-view, this may not be feasible. The third method is based on simple analytical calculations (e.g., \citealt{newmanDavis, Driver_2003, Somerville_2004, Trenti_2008, moster}) and is hence widely adopted in the literature. However, this method depends on two simplifying assumptions about the survey geometry and the underlying fluctuations, and it can lead to a significant underestimate of the sample variance in a survey with a narrow field-of-view. Here we present a new method to improve the standard analytical calculations without any restriction to the simplifying assumptions in the standard method.

As an application of our new method, we consider cosmological observations with a narrow field-of-view to find the missing baryons in the local Universe (\citealt{Fukugita_1998}; see \citealt{Bregman_2007} for a recent review). The baryon density in the Universe can be observationally estimated by inferring the baryon mass contained in stars, hot/cold gas, and the intergalactic medium (see \citealt{Fukugita_1998, Shull_2012} and the references therein). These estimates in the local neighborhood, however, fall short by about 30\%, compared to the baryon density parameter~$\Omega_b = 0.049$ inferred from Big Bang nucleosynthesis and CMB observations, while the amount of baryons found from Lyman-$\alpha$ forests at $z=2-4$ is consistent with $\Omega_b = 0.049$ (see, e.g., \citealt{Weinberg_1997, PITROU20181, planck18}). A large amount of theoretical and numerical work (\citealt{Dave_2001, Fang_2002, Maller_2004, Fang_2013}, \citealt{Shull_2012, Driver_2021}) has been performed to show that the missing baryons in the local Universe are shock heated, residing in the WHIM, and X-ray or far UV observations are needed to confirm their presence.

In 2005 and 2018, X-ray spectra of blazars were obtained to measure the intervening absorption lines from highly ionized oxygen contained in the WHIM by using the Low Energy Transmission Grating (LETG) on Chandra \citep{Nicastro_2005} and the Reflection Grating Spectrometer (RGS) on the X-ray Multi-Mirror (XMM-Newton) mission \citep{nicastro}. We will refer to these two observations as the 2005 and 2018 WHIM observations, respectively. The estimates of the ionized oxygen column densities are then used to infer the amount of baryons in the WHIM, following the method of \cite{Savage_2002}. Both observations found that the WHIM contains the right amount of baryons missing in the local neighborhood (see, e.g., \citealt{Dai_2010, Kovacs_2019} for other observations). With observations along two line-of-sights, their estimates of the cosmic mean value of baryons in the WHIM are naturally subject to large sample variance errors. Here we compute the sample variance for these surveys and find the sample variance errors are in fact larger than the reported measurement uncertainties. When estimating the sample variance, the blazars we consider are in different regions of the sky so we do not attempt to stack observations like in \cite{Kovacs_2019}. Fast radio burst observations in the local neighborhood offer an estimate of the baryon density parameter consistent with $\Omega_b = 0.049$ \citep{Macquart_2020}, but since they are only observed along a few line-of-sights, these observations are also subject to large sample variance errors.

For comparison and reference, we ran ten dark matter-only simulations using GADGET-2 \citep{gadget}. Our simulations were set up with $512^3$ particles and $16$~$h^{-1}$kpc softening lengths in a $300$~$h^{-1}$Mpc periodic box by using 2LPTic code \citep{2lpt} at~$z=49$. The transfer function was computed using the Boltzmann solver CLASS \citep{class}, and we adopt the $\Lambda$CDM parameters from \cite{planck18}: Hubble constant~$h = 0.67$, baryon density parameter~$\Omega_{b}=0.049$, dark matter density parameter~$\Omega_{\text{DM}}=0.27$, the spectral index~$n_{s}=0.97$, and its primordial fluctuation amplitude~$\ln(10^{10}A_s) = 3.0$.
\vspace{0.5cm}

\section{Analytical Modeling of Sample Variance}\label{sec:methods}
Here we present our analytical method for modeling the sample variance of cosmological observations in a given survey, characterized by the opening angle~$\alpha$ and the redshift range~$[z_{\text{min}},z_{\text{max}}]$. Measurements of distant sources in the survey yield an observable~${\mathcal O}(\lambda,\hat n)$ such as the flux of a source, where $\lambda$ is the wavelength of the measurements and $\hat n$ is the angular position of the source within the survey. The main observable is often constructed by averaging~${\mathcal O}$ over the angular position or integrating~$\mathcal O$ over time, but this mean value of the observable has the fluctuation around the true value, due to the intrinsic stochasticity. Splitting the observable~$\mathcal O \equiv \overline{\mathcal O} + \delta \mathcal O$ into the background~$\overline{\mathcal O}$ and the fluctuation~$\delta \mathcal O$, the (dimensionless) sample variance can be written \citep{PeeblesLSS_1980} as
\begin{eqnarray}\label{eqn:cv6d}
    \sigma^2\equiv \left\langle \left(\frac{\delta \mathcal O}{\overline{\mathcal O}} \right)^2 \right\rangle
    = \frac{1}{V^2} \int dV_a dV_b ~\xi_{\delta \mathcal O}(r_{ab}) ~,
\end{eqnarray}
where $V$ is the survey volume, $\xi_{\delta \mathcal O}$ is the two-point correlation function of the observable~$\mathcal O$, and $r_{ab}$ is the separation between any two points inside the survey. The sample variance arises from the fluctuation in $\mathcal O$ and its correlation over the survey volume~$V$.

\begin{deluxetable*}{cccccccc}[t]\label{tab:surveys}
    \tablecaption{Survey redshift and geometries for a given observation}
    \tablecolumns{8}
    \tablehead{\colhead{} & \multicolumn{3}{c}{2005 Observation} & \multicolumn{3}{c}{2018 Observation} \\ \cline{2-4}\cline{5-7}
    \colhead{Geometry} & \colhead{$z_{\text{min}}$} & \colhead{$z_{\text{max}}$} & \colhead{Vol. coef. $C$} & \colhead{$z_{\text{min}}$} & \colhead{$z_{\text{max}}$} & \colhead{Vol. coef. $C$}}
    \startdata
    Line-of-sight & 0 & 0.0308 (Blazar) &  & 0 & 0.49 (Blazar) &  \\
    Cone & 0 & 0.0308 (Blazar) & 0.0171 & 0 & 0.49 (Blazar) & 47.8 \\
    Box 1 & 0 & 0.0308 (Blazar) & 0.0164 & 0 & 0.49 (Blazar) & 52.1 \\
    Box 2 & 0 & 0.027 (Absorber 2) & 0.0111 & 0 & 0.4339 (Absorber 2) & 37.3 \\
    Box 3 & 0.011 (Absorber 1) & 0.0308 (Blazar) & 0.0193 & 0.3551 (Absorber 1) & 0.49 (Blazar) & 34.9 \\
    Box 4 & 0.011 (Absorber 1) & 0.027 (Absorber 2) & 0.0129 & 0.3551 (Absorber 1) & 0.4339 (Absorber 2) & 18.4 
    \enddata
    \tablecomments{Survey volume is $V = C \left(\alpha / \text{arcmin}\right)^2$ $h^{-3}$Mpc$^3$, where the opening angle~$\alpha$ is a free parameter.}
\end{deluxetable*}
\begin{figure*}[t]
\vspace{-0.8cm}
		{\centering	\includegraphics[width=0.49\linewidth]{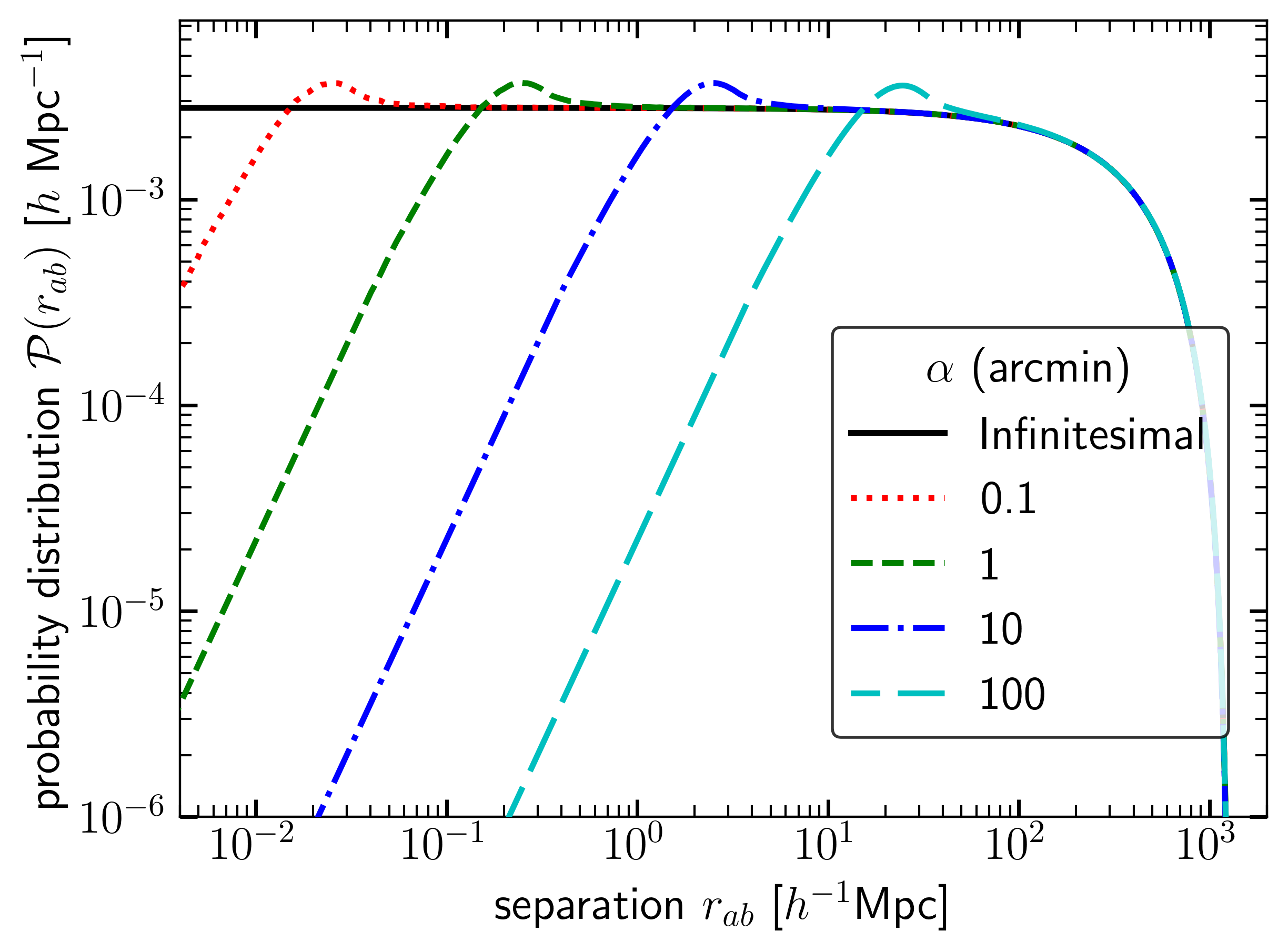}
        \includegraphics[width=0.49\linewidth]{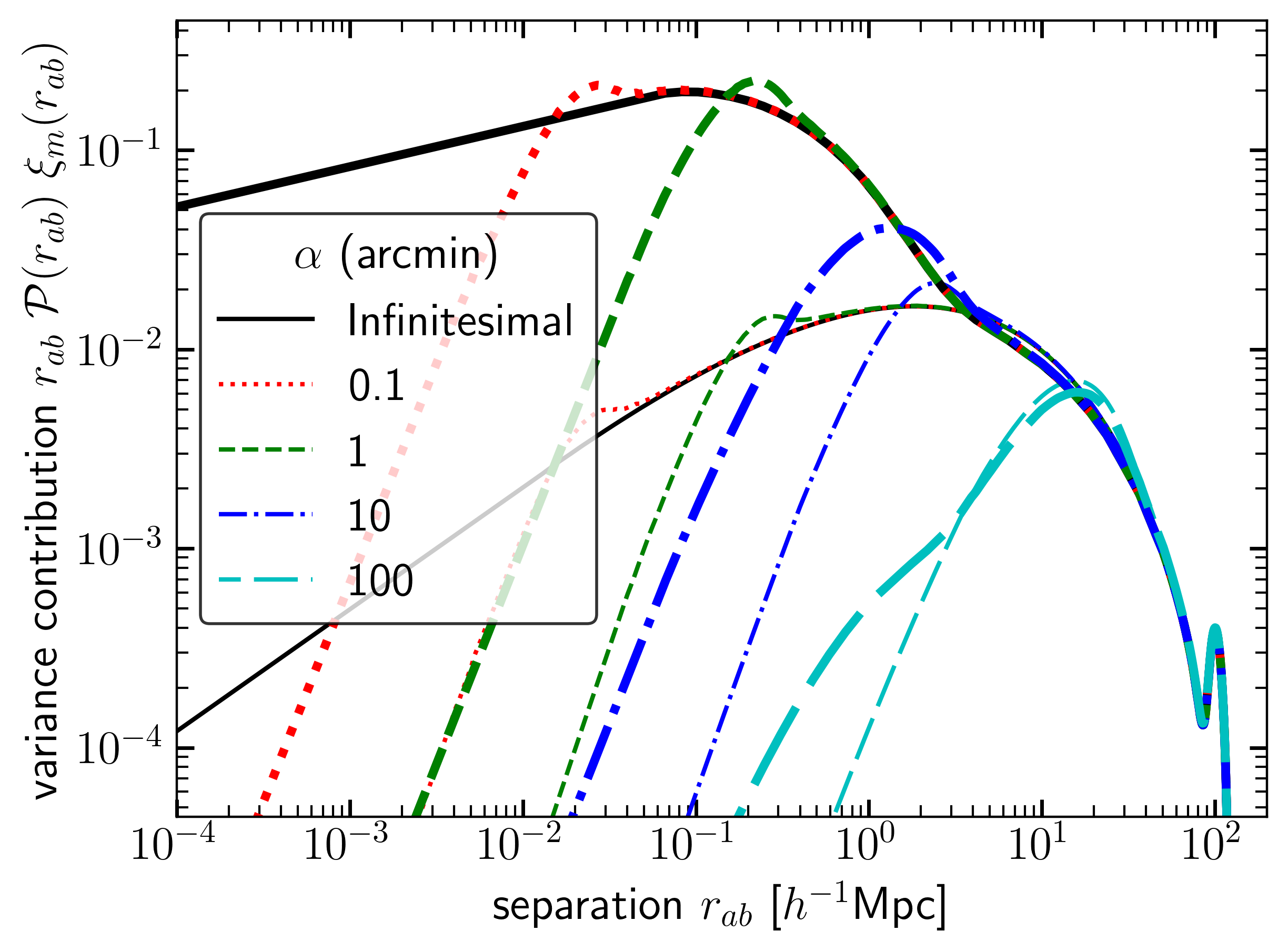}
		\caption{Probability distribution of separations~$r_{ab}$ (left) and contributions to the sample variance (right) for cone-shaped surveys in the~2018 observation. We consider various opening angles~$\alpha$. In the left panel, we plot the probability distributions for infinitesimal~$\alpha \rightarrow 0$ and then four logarithmically spaced values from $\alpha=0.1$ to $\alpha=100$. The left panel determines the effect of the survey geometry on the sample variance for a given separation~$r_{ab}$ and opening angle~$\alpha$. The right panel shows the dark matter two-point correlation function~$\xi_m$, either the nonlinear case (thick) or linear case (thin), weighted by the probability distribution. Integration of these curves yields the sample variance according to Equation~(\ref{eqn:sigmaR1d}).
        }
		\label{fig:pdfAndConvolution}
		}
\end{figure*}
\vspace{-0.85cm}

The standard analytical method in the literature is outlined in \citet{moster}, which computes the sample variance $\sigma_m$ in Equation~(\ref{eqn:cv6d}) from dark matter fluctuations
\begin{equation}\label{eqn:cvStandard}
    \sigma_m^2 = \int \frac{d^3k}{(2\pi)^3} ~ P_m(k)W^2(\bm k) ~,
\end{equation}
where $P_m(k)$ is the linear matter power spectrum and $W(\bm k)$ is the survey window function. In deriving Equation~(\ref{eqn:cvStandard}), we assumed that the observable is a dark matter fluctuation~$\delta \mathcal O = \delta_m$, its two-point correlation function is just a function of separation~$r_{ab}$, and the survey geometry is a rectangular box, whose depth~$\Delta L_z$ is set by the redshift range, and width~$\Delta L$ is determined at the mean redshift by the opening angle, such that the window function is $W(\bm k) = 
j_0 \left(k_x \Delta L/2\right) 
j_0 \left(k_y \Delta L/2\right)
j_0 \left(k_z \Delta L_z/2\right)$,
where $j_0$ is a spherical Bessel function. 

In this work, we will adopt the same assumptions as in the standard method, but improve on two aspects in computing the sample variance: We will use the nonlinear two-point correlation function and account for the correct geometry of a given survey. Here we model the survey geometry as a light cone without any holes within the boundary set by the opening angle~$\alpha$. To facilitate the computation of the sample variance in a given geometry, we rewrite Equation~(\ref{eqn:cv6d}) as
\begin{equation}\label{eqn:sigmaR1d}
    \sigma^2 = \int dr_{ab} ~\xi_{\delta \mathcal O}(r_{ab}) \mathcal P(r_{ab}) ~,
\end{equation}
in terms of the probability distribution~$\mathcal P$ of a pair separation~$r_{ab}$ in a given geometry. The difficulty in the volume integral over a given geometry in Equation~(\ref{eqn:cv6d}) is now replaced with finding the probability distribution~$\mathcal P(r_{ab})$ in Equation~(\ref{eqn:sigmaR1d}), while the one-dimensional integral is far simpler to perform than the six-dimensional volume integral.  

Computation of the sample variance with our method needs two ingredients. First, our $N$-body simulations are used to compute the nonlinear two-point correlation function of dark matter fluctuations. Second, the probability distribution~$\mathcal P(r_{ab})$ is obtained by a Monte Carlo method in a given geometry unless an exact solution is known for the probability distribution, as in the case of a sphere (see, e.g., \citealt{tu}) or a box (J. Philip 2007). In particular, we are interested in the geometry for describing the spectroscopic measurements of blazars along the line-of-sight direction, in which the probability distribution can be readily derived as
\begin{equation}\label{eqn:pdfLOS}
    \mathcal P(r_{ab})  
    = -\frac{3 r_{ab}^5}{5 R^6} + \frac{6 r_{ab}^2}{R^3} - \frac{9 r_{ab}}{R^2} + \frac{18}{5 R}~, 
\end{equation}
where $R$ is the maximum pair separation. The line-of-sight we consider is constructed by taking a cone geometry, which has a uniform distribution of points inside, and taking the limit that the cone's opening angle becomes infinitesimal. 

For simplicity, we assume that the WHIM observable~$\mathcal O$ traces the dark matter fluctuation exactly such that the distribution of WHIM in a survey matches the distribution of dark matter and they both have the same two-point correlation function. We also used the  \citet{Giri_2021} emulator to model the total matter fluctuation including baryonic effects. Furthermore, as pointed out in \citet{jyooBaryon}, the observer position is not a random position, but a special position (i.e., Milky Way halo), such that it contributes to the sample variance in terms of additional two-point and three-point correlations with one point anchored at the observer position. However, we verified from our $N$-body simulations that these extra contributions missing in the literature are negligible for the low-redshift surveys considered in this work.

\section{Application to the WHIM observations}\label{sec:results}
We are now ready to apply our analytical method to the missing baryon problem in the local neighborhood. To model the geometry of the 2005 and 2018 observations \citep{Nicastro_2005, nicastro}, we treat each survey as a narrow cone with the observer at the vertex, extending to the distant background source, and the WHIM absorbers are contained within the volume of the cone. Each observation has found two WHIM absorbers against the background blazars at $z=0.0308$ and $z=0.49$, respectively. In contrast, the standard procedure in the literature is to approximate the survey geometry as a box, instead of a cone, with various boundaries in redshift. We consider six different geometries for each observation, and their redshift ranges are listed in Table~\ref{tab:surveys}. All six geometries share the common opening angle~$\alpha$ that we vary with values comparable to the field-of-view of the XMM-Newton RGS, which is $5$~arcminutes. Since the WHIMs are detected by spectroscopic observations of the blazars, the true geometry is close to the cone in the limit~$\alpha \rightarrow 0$, which is referred to as \textit{line-of-sight} geometry in Table~\ref{tab:surveys}.   
\begin{figure}[t]
		{\centering	\includegraphics[width=\linewidth]{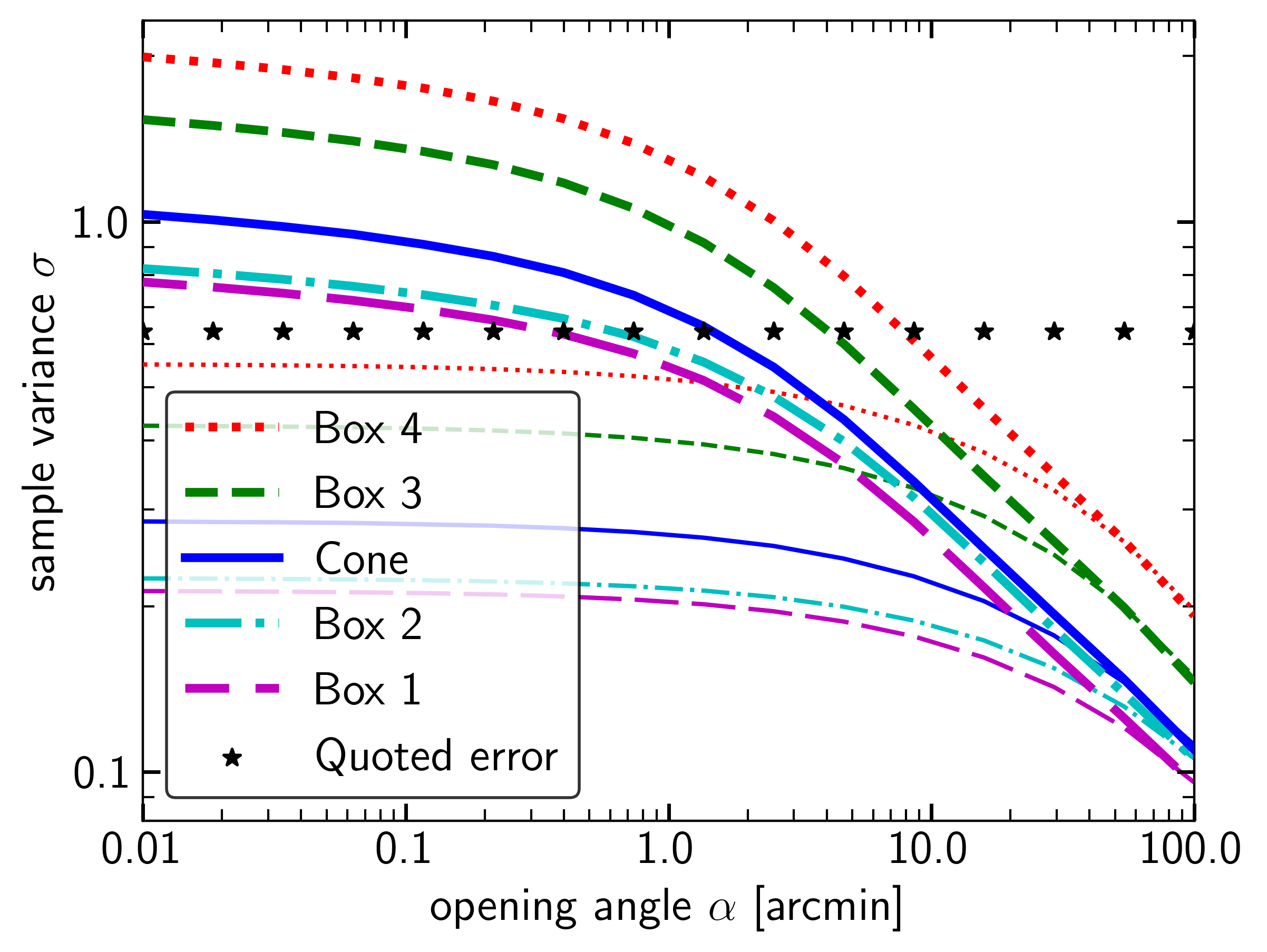}
		\caption{Sample variance~$\sigma$ for five different survey geometries computed with a nonlinear (thick) or linear (thin) dark matter two-point correlation function for the 2018 observation as a function of opening angle~$\alpha$. In the limit~$\alpha \rightarrow 0$, the cone geometry reduces to the line-of-sight geometry. The horizontal stars, which do not depend on $\alpha$, are the 2018 observation error quoted in \cite{nicastro}.
		}
		\label{fig:sv}
		}
\vspace{-0.3cm}
\end{figure}

The left panel in Figure~\ref{fig:pdfAndConvolution} shows the probability distribution of pair separations~$r_{ab}$ for the 2018 observation described by the cone geometry in Table~1, as a function of opening angle~$\alpha$. At a given separation~$r_{ab}$, the probability distribution is fully determined by the survey geometry. Let $r_*$ be the cone's base radius at the maximum redshift of the survey, which is approximately located at the peak of the cone's probability distribution. For large pair separations~$r_{ab} \gg r_*$, only pairs separated along the line-of-sight direction can be contained in the geometry. Therefore, the probability distributions behave in the same way independent of the opening angle~$\alpha$. For the opposite case of small pair separations~$r_{ab} \ll r_*$, the survey geometry becomes irrelevant, and the probability distributions exhibit a self-similar behavior with $r_{ab}^2$ scaling due to pairs of points being uniformly distributed with a probability distribution that is proportional to the surface area of a sphere. The probability distribution for box-shaped surveys in Table~\ref{tab:surveys} exhibits a similar behavior upon replacing $r_*$ with the box's side length tangential to the line-of-sight. The right panel of Figure~\ref{fig:pdfAndConvolution} shows the contribution to the sample variance: the two-point correlation weighted by the probability distribution as a function of $r_{ab}$. According to Equation~(\ref{eqn:sigmaR1d}), it is the two-point correlation weighted by the probability distribution, not the probability distribution alone, that contributes to the sample variance in a given survey. The thick and thin curves show contributions to the sample variance obtained by using the nonlinear dark matter two-point correlation function~$\xi_{\text{NL}}$ from $N$-body simulations and the linear matter correlation function~$\xi_{\text{lin}}$ from CLASS, respectively. Given that $\xi_{\text{NL}}$ deviates from $\xi_{\text{lin}}$ on small scales~$r_{ab} < 5 $~$h^{-1}$Mpc and the probability distribution of pair separations~$r_{ab}$ is determined by $r_*$ (or~$\alpha$), the difference in contributions to the sample variance is negligible for surveys with large opening angles, but is especially pronounced for surveys with small opening angles. Indeed, the boost in the sample variance is quite dramatic for narrow surveys ($\alpha < 1$~degree), which illustrates the significance of the nonlinearity and survey geometry in computing the sample variance in narrow surveys. 

Figure~\ref{fig:sv} plots the sample variance for five different survey geometries considered for the 2018 observation. Again, thick and thin curves represent the sample variance computed by using the nonlinear dark matter correlation function~$\xi_{\text{NL}}$ and the linear dark matter correlation function~$\xi_{\text{lin}}$, respectively. In general, the sample variance in all cases increases as the survey volume decreases (or $\alpha$ decreases). However, the difference in the sample variance between the thick and thin curves arises, not only due to the differences in $\xi_{\text{NL}}$ and $\xi_{\text{lin}}$, but also due to the difference in the probability distributions for pair separations, as shown in Figure~\ref{fig:pdfAndConvolution}.

The standard method for estimating the sample variance approximates the survey geometry as a box, whose lower and upper bounds of the WHIM observations are rather ill-defined, as opposed to galaxy redshift surveys for example. Four different cases of the box geometry in Table~\ref{tab:surveys} represent the standard method in which the base length is fixed by the opening angle at the average redshift, as described in Section~\ref{sec:methods}. For the 2018 observation, the standard method predicts $\sigma=0.20$ (long-dashed thin line) for Box~1 in Table~\ref{tab:surveys} at $\alpha=1$~arcmin while a more realistic cone model predicts $\sigma=0.69$ (solid thick line). Note that the survey volumes for Box~1 and the cone geometries are roughly equal. If $\xi_{\text{NL}}$ is adopted for the standard method, it predicts $\sigma=0.55$ (long-dashed thick line). For a fixed opening angle and fixed two-point correlation function model, Boxes 2, 3, and 4, respectively, have a sample variance that is increasingly greater than the sample variance for Box 1, mostly due to their decreasing volumes. For our best model, ``line-of-sight" geometry, the sample variance increases to $\sigma=1.19$ (solid thick line at $\alpha=0$). 

As given in \citet{nicastro}, the amount of baryons contained in the WHIM is between $9\%$ to $40\%$ of $\Omega_b$. If we rewrite this range as approximately $\Omega_{\text{WHIM}} / \Omega_b = 0.245~(1 \pm \sigma)$, then their quoted error corresponds to the dimensionless error~$\sigma=0.63$ (horizontal stars). In the same way, we analyzed the 2005 observation. Our estimate of the sample variance for the 2005 observation is again larger than the quoted error, and the discrepancy in the sample variance estimates between the standard method and our method is also stronger, as the survey volume for the 2005 observation is smaller compared to the 2018 observation. Since the overall trends are similar for the 2005 observation, we do not include additional plots like Figures~1~and~2.

Assuming that our model with the line-of-sight geometry, obtained by taking a cone geometry in the limit of opening angle~$\alpha \rightarrow 0$, best describes the WHIM observations, we estimate the sample variance as $\sigma=1.19$ and $\sigma=4.45$ for the 2018 and 2005 observations, respectively. These numbers are significantly larger than the observational errors~$\sigma=0.63$ and $\sigma=0.78$ found in \citet{nicastro} and \citet{Nicastro_2005}, respectively. The standard method described as the box cases in Table~\ref{tab:surveys} fails to capture the correct geometry and, if used with a linear two-point correlation function, significantly underpredicts the sample variance as shown in Figure~\ref{fig:sv}. Our estimates of the sample variance change very little when we use the total matter fluctuations, including baryonic effects \citep{Giri_2021}. Until now, we have approximated the two-point correlation using the matter distribution at $z=0$. If we instead account for cosmic evolution by multiplying Equation~
\ref{eqn:sigmaR1d} by the linear growth factor~$D(z)$ at the survey's mean redshift, the sample variance decreases by 12\% and less than 1\% for the 2018 and 2005 observations, respectively.
\vspace{0.1cm}

\section{Conclusion and Discussion}\label{sec:conclusion}
In this Letter, we have computed the sample variance of a survey with a small field-of-view, accounting for the nonlinearity of the underlying fluctuations and the correct geometry of the survey. Our method is based on Monte Carlo simulations and is readily applicable to any survey geometry. In contrast, the standard method for computing the sample variance ignores the nonlinearity on small scales and approximates the survey geometry as a box. Compared to the standard method, we have found that these two factors play a dramatic role in computing the sample variance and the correct sample variance is greatly underestimated in a survey with a small field-of-view such as spectroscopic observations of the intergalactic medium against remote sources. 

To demonstrate the impact on interpreting observations, we have applied our method to the 2005 and 2018 WHIM observations (\citealt{Nicastro_2005, nicastro}) and found that the sample variance is indeed larger by a factor two than the quoted error for the 2018 observation (and close to a factor six for the 2005 observation). Our estimates of the sample variance are based on the critical assumption that the WHIM fluctuates around the background in the same way as dark matter fluctuates (i.e., a linear bias~$b=1$). Though the WHIM is expected to trace dark matter on large scales similarly to baryons, the exact relation between the WHIM and dark matter distributions on small scales is certainly more complicated than assumed in this work.

With large sample variances in two observations, the amount of baryons found in the WHIM observations is statistically unlikely to represent the cosmic mean value for baryons contained in the WHIM. For example, a $1\text{-}\sigma$ fluctuation present in the WHIM can match the amount found in the 2018 observation with the cosmic mean density~$\Omega_{\text{WHIM}} / \Omega_b = 0.112,$ lower by $54\%$ than the estimated amount in the 2018 observation. The 2005 observation has a larger sample variance than the 2018 observation and can match a lower cosmic mean density. Though these observations are critical in solving the missing baryon problem, it is indeed challenging to derive proper interpretations of cosmological observations in surveys with a narrow geometry, and our application demonstrates the impact of accurately quantifying the sample variance.
\\
 
This work is supported by the Swiss National Science Foundation Grant CRSII5\_173716.

\bibliographystyle{aasjournal.bst}
\bibliography{sampleVariance.bib}

\end{document}